\documentclass[secnumarabic,amssymb,amsmath,natbib,graphicx,aps,twocolumn]{revtex4-2}
\usepackage{lipsum}
\usepackage{subcaption}
\usepackage{tikz}       
\usepackage{xcolor} 
\usepackage[colorlinks=true,citecolor=blue,hidelinks]{hyperref}
\usepackage{comment}
\usepackage{booktabs}
\usepackage{float}
\usepackage{dcolumn}

\def\bea{\begin{eqnarray}}
\def\eea{\end{eqnarray}}

\begin{document}

\title{Probing the Macroscopic Imprints of Singularity Resolution: Polarization Diagnostics of Rotating ABG Black Holes}
\author{Ziyang Li}
\author{Jia-Hui Huang}
\email{huangjh@m.scnu.edu.cn}
\affiliation{Key Laboratory of Atomic and Subatomic Structure and Quantum Control (Ministry of Education), Guangdong Basic Research Center of Excellence for Structure and Fundamental Interactions of Matter, School of Physics, South China Normal University, Guangzhou 510006, China}
\affiliation{Guangdong Provincial Key Laboratory of Quantum Engineering and Quantum Materials, Guangdong-Hong Kong Joint Laboratory of Quantum Matter, South China Normal University, Guangzhou 510006, China}

\date{\today}

	\begin{abstract}
	Recent observations by the Event Horizon Telescope have revealed important electromagnetic information around the supermassive black holes $\text{M87}^*$ and $\text{Sgr A}^*$, providing experimental supports for investigating physics near the event horizon. In the absence of a complete quantum gravity theory, exploring the observational appearances and polarized images of accretion flows in regularized geometries, and seeking specific observational differences from Kerr black holes, is of value for understanding the singularity problem. Based on backward ray tracing method, this work investigates both the optical and polarimetric observational features of rotating Ay\'{o}n-Beato-Garc\'{i}a spacetime with regularization parameter $Q$. It is found that, compared with the Kerr case, the parameter $Q$ leads to a significant contraction of the size of the inner shadow in the observed optical images. This contraction remains visible even after an illustrative Gaussian filter for images observed at a low inclination angle (e.g. $17^\circ$).
In the polarized images, we focus on two observables, the electric vector position angles (EVPAs) and the depolarization regions, on the observer's screen. 
It is found that depolarization regions do not necessarily reside in the vicinity of the bright photon ring in images at high inclination angles (e.g. $83^\circ$), and a multi-branch structure can be observed. Furthermore, the distribution of EVPAs is insensitive to the regularization parameter $Q$ while the morphology of the depolarization regions can be impacted remarkably by $Q$. The results provide possible optical observable characteristics for the rotating regular spacetime.
	\end{abstract}

\maketitle	

	\section{Introduction}
	The observation of gravitational waves by the LIGO/Virgo collaboration \cite{Abbott:2016blz} and the horizon-scale imaging of the compact objects M87* \cite{EventHorizonTelescope:2019dse} and Sgr A* \cite{EventHorizonTelescope:2022wkp} provide strong evidence for astrophysical black holes and enable tests that are consistent with general relativity (GR) in the strong-gravity regime. However, GR predicts spacetime singularities \cite{Penrose:1965,Hawking:1970}, where classical GR breaks down \cite{Marolf:2017jkr}. Regular black-hole models were introduced as one possible framework for replacing the singular region by a regular core \cite{Bardeen1968, Hayward:2006part,Bambi:2023try,Lan:2023cvz,Zhang:2025ccx}. However, regularity of curvature invariants and geodesic completeness should be assessed separately \cite{Guerrero:2023unv}.
	
Over the past few years, optical imaging features of black holes have emerged as a powerful probe for testing GR and modified gravity theories in the strong-field regime \cite{EventHorizonTelescope:2020qrl,Vagnozzi:2022moj,Allahyari:2019jqz,Chen:2022scf,Gao:2023mjb,Zeng:2023zlf,Zeng:2022pvb,Zeng:2021mok,Li:2025cnv,Xiong:2025hjn,Li:2024ctu,Cao:2024vtq,
Li:2024kyv,Chen:2025ujl,Cao:2025qzy,Wang:2025edt,
Zhao:2025ouq,Wang:2025vsx,Zeng:2025tji,He:2026fyz,Zeng:2025kqw,Li:2026vpa,Liu:2025hhg,Shu:2024tut, Ye:2023qks}, such as those predicting regular black hole models \cite{Bambi:2013ufa,Toshmatov:2014nya,Mazza:2021rgq,Simpson:2018tsi,Bronnikov:2021uta,Johannsen:2013szh,Fan:2016hvf}, aiming to constrain theoretical model parameters or search for potential observational evidence of deviations from GR through observational signatures. 
	
Among the various nonsingular spacetimes, Ayon-Beato and Garcia (ABG) proposed a novel class of black bounce geometries \cite{Ayon-Beato:1998hmi}, which are obtained by coupling the gravity to some nonlinear electromagnetic sources and provide an explanation to the Bardeen black hole \cite{Ayon-Beato:2000mjt}.
 From an astrophysical perspective, since realistic compact objects are expected to possess angular momentum \cite{Kerr:1963ud}, the static regular ABG black hole solution was generalized to the rotating case \cite{Toshmatov:2014nya}. This rotating ABG metric may serve as an Kerr mimicker while resolving the spacetime singularity, and provides an ideal laboratory for testing strong-field gravity.

The spherical and rotating ABG metrics have been studied from various aspects recently in the literature, such as thermodynamics of ABG spacetime \cite{Hegde:2021gfs}, the strong gravitational lensing and shadows \cite{Ramadhan:2023ogm,Fauzi:2025asu}, the chaotic behaviors of particle motion in the ABG black hole coupled to a cloud of strings \cite{Liu:2026npn}, and quasinormal modes and grey-body factors \cite{Maji:2026fwa}. Particularly, the thermal disk properties, observational intensity images and redshift factors of the rotating ABG black holes have been discussed in \cite{Ban:2026ciz}.

In this work, we perform a study of the polarized images of the rotating ABG black holes surrounded by a thin accretion disk. We consider two representative magnetic configurations around the black hole and focus on two observables, the distribution of electric vector position angles (EVPAs) and the depolarization regions.
This paper is organized as follows.
In Sec.~\ref{sec:model}, we introduce the rotating ABG black hole metric, the accretion disk model and the magnetic field configurations. 
In Sec.~\ref{sec:results}, we present and discuss the numerical results of the polarized images for various model parameters and inclination angles. For convenience of discussion, we also present the corresponding observed intensity images of the black hole. 
 Sec.~\ref{sec:Conclusion} is devoted to the conclusion.
 
\section{The Model and Setup}
\label{sec:model}

\subsection{Spacetime Geometry and Photon Dynamics}
\label{subsec:spacetime}

We consider the Newman-Janis-generated rotating extension of the ABG geometry \cite{Toshmatov:2014nya}. Its static seed is sourced by nonlinear electrodynamics (NLED), and the rotating metric asymptotically approaches the Kerr-Newman form in the far-field limit. In geometric units with $G=c=1$, the line element of the rotating ABG geometry in Boyer-Lindquist coordinates $(t, r, \theta, \phi)$ is given by
\begin{equation}
	\label{eq:metric}
	\begin{aligned}
		ds^2 = & -f dt^2 - 2a \sin^2\theta (1 - f) dt d\phi + \frac{\rho^2}{\Delta} dr^2 \\
		& + \rho^2 d\theta^2 + \sin^2\theta \left[ \rho^2 - a^2 (f - 2) \sin^2\theta \right] d\phi^2,
	\end{aligned}
\end{equation}
where
\begin{align}
	\rho^{2} &= r^{2}+a^{2}\cos^{2}\theta, \\
	\Delta &= \rho^{2}f+a^{2}\sin^{2}\theta, \\
	f &= 1-\frac{2Mr\rho}{(\rho^{2}+Q^{2})^{3/2}}+\frac{Q^{2}\rho^{2}}{(\rho^{2}+Q^{2})^{2}}.
\end{align}
Here, $M$ denotes the black hole mass ($M=1$ is adopted hereafter), $a$ represents the black hole spin parameter, and $Q$ is the NLED charge parameter and is also the regularization parameter of the spacetime core. In the limit $Q\rightarrow0$, the metric recovers the standard Kerr solution. The event horizon is determined by the largest root of $\Delta(r,\theta)=0$~\cite{Toshmatov:2014nya}. 

The observed intensity and polarized images can be obtained with the backward ray-tracing method by tracking null geodesics in this curved geometry. The photon trajectories are governed by the canonical Hamiltonian equations
\begin{equation}
	\dot{x}^{\mu}=\frac{\partial\mathcal{H}}{\partial p_{\mu}}, \quad \dot{p}_{\mu}=-\frac{\partial\mathcal{H}}{\partial x^{\mu}},
\end{equation}
subject to the following constraint for massless particles
\begin{equation}
	\mathcal{H}=\frac{1}{2}g^{\mu\nu}p_{\mu}p_{\nu}=0.
\end{equation}
The resulting trajectories encode the lensing structure of the spacetime and determine the formation of the central shadow and the bright photon ring features in the optical images. By integrating these equations, photons propagate from the observation screen to the emission region in the accretion disk, providing the foundation for constructing both intensity and polarized images. Throughout this work, the observed radiation is treated as a test field following null geodesics of the background metric. Possible modifications associated with a NLED effective optical metric are not included.

\subsection{Equatorial Thin Disk and Emission Model}

Since we focus on the effects of spacetime geometry on polarimetric images, we adopt an optically and geometrically thin accretion disk in the equatorial plane as the emission model. This choice minimizes radiative transfer effects (e.g., Faraday rotation, multiple scattering) in the disk and allows us to isolate purely geometric modifications to the polarization pattern \cite{Chael:2021rjo}.
We also assume that the emitting plasma moves on equatorial circular orbits outside the innermost stable circular orbit (ISCO). The angular velocity is given by
\begin{equation}
	\Omega_{\pm}
	=
	\frac{-\partial_r g_{t\phi}
		\pm
		\sqrt{(\partial_r g_{t\phi})^2
			-
			(\partial_r g_{tt})(\partial_r g_{\phi\phi})}}
	{\partial_r g_{\phi\phi}},
\end{equation}
where the plus (minus) sign corresponds to prograde (retrograde) motion. Unless otherwise specified, we consider prograde motion in this work.
The ISCO radius separates stable circular motion from the plunging region and is determined numerically from the radial effective potential. Outside the ISCO, the plasma follows circular motion with four-velocity $u^\mu = (u^t,0,0,u^\phi)$, while inside the ISCO it forms a plunging flow. In the plunging region, we assume that the conserved quantities are inherited from their ISCO values, which uniquely determines the velocity field. The four-velocity directly enters the redshift factor through the scalar product $P_\mu u^\mu$, thereby governing the observed intensity distribution.
This minimal disk model fixes the phenomenological emission prescription across the parameter survey. The resulting differences therefore represent the response of this controlled toy model to the spacetime parameters, rather than a general removal of plasma-model degeneracies.

Motivated by general relativistic magnetohydrodynamic (GRMHD) simulations, which show that the time-averaged emissivity is strongly concentrated near the equatorial plane, we further adopt a simplified analytical emission model for the geometrically thin and optically thin disk. The emission is assumed to be isotropic in the rest frame of the plasma, with both the emissivity and the velocity field depending only on the radial coordinate.
Within this framework, the observed specific intensity can be expressed as
\begin{equation}
	I_{\rm obs} = \sum_{n=1}^{N_{\max}} f_n\, g_n^3\, j(r_n),
\end{equation}
where $g_n$ is the redshift factor associated with the $n$-th disk crossing and $j(r_n)$ is the emitted specific-intensity profile evaluated at the corresponding crossing radius. 
We denote the contribution of this crossing by $I_n\equiv f_n g_n^3j(r_n)$. The cubic dependence on $g_n$ follows from the invariance of $I_\nu/\nu^3$ along a vacuum null ray \cite{Chael:2021rjo}. A light ray may cross the accretion disk multiple times due to strong gravitational lensing, and each crossing contributes to the total observed intensity. Since higher-order contributions are highly suppressed in this toy model, we truncate the sum at $N_{\max}=3$, which captures contributions from the direct image and the first two disk crossings. We adopt the phenomenological weighting factors $f_1=1$ and $f_{2,3}=2/3$ \cite{Chael:2021rjo}.
The redshift factor is defined as
\begin{equation}
	g = \frac{(P_\mu u^\mu)_{\rm obs}}{(P_\mu u^\mu)_{\rm em}},
\end{equation}
where $P^\mu$ is the photon four-momentum and $u^\mu$ is the four-velocity of the emitter or observer. This factor encodes both gravitational redshift and Doppler effects.
For the emissivity, we adopt a phenomenological radial profile given by
\begin{equation}
	j(r_n) = \exp\left(p_1 z + p_2 z^2\right), \qquad
	z = \log\left(\frac{r_n}{r_h}\right),
\end{equation}
where $r_h$ denotes the radius of the event horizon. The parameters $(p_1, p_2)$ control the radial distribution of emission and are chosen to reproduce representative profiles inferred from GRMHD simulations. In particular, we consider $(p_1, p_2)=(-2,-1/2)$ as a fiducial choice motivated by 230~GHz imaging models \cite{Chael:2021rjo}.

\subsection{Magnetic Field Configuration and Polarization}
\label{sec:polarization}

Synchrotron radiation from magnetized plasma produces linearly polarized emission, making polarization a powerful probe of the near-horizon region. EHT observations have revealed spatially resolved polarization structure in M87* \cite{EHT_M87_pol}, motivating controlled models of polarized equatorial emission \cite{Gelles:2021}. A central objective of this study is to study how the regular ABG spacetime shape the polarization patterns. 

In the fluid rest frame of the emitting plasma, the orientation of optically thin synchrotron polarization is perpendicular to the projected magnetic field \cite{Gelles:2021}. At the emission point, the spatial polarization direction $\vec{f}$ can be expressed as
\begin{equation}
	\label{eq:f_spatial}
	\vec{f} = \frac{\vec{k} \times \vec{B}}{|\vec{k} \times \vec{B}|}.
\end{equation}
To consistently trace the polarization transport across the curved geometry, this spatial geometric relation must be elevated to a generally covariant four-dimensional tensor framework. With the Levi-Civita tensor $\epsilon^{\mu\nu\rho\sigma}$ in curved spacetime, the generalized covariant representation for the polarization four vector $f^\mu$ is rigorously formulated as
\begin{equation}
	\label{eq:f_covariant}
	f^\mu = \frac{1}{\mathcal{N}} \epsilon^{\mu\nu\rho\sigma} u_\nu k_\rho B_\sigma,
\end{equation}
where $\mathcal{N}$ serves as a scalar normalization factor. The polarization vector satisfies:
\begin{equation}
	\label{eq:f_normalization}
	f^\mu f_\mu = 1.
\end{equation}
In these expressions, $\vec{k}$ and $\vec{B}$ denote the spatial three-momentum of the photon and the local magnetic field vector evaluated in the plasma rest frame, respectively. The normalized expression is well defined when the pitch angle is nonzero; in a complete synchrotron model the emissivity vanishes when $\vec{k}\parallel\vec{B}$. Additionally, $u_\nu$ corresponds to the covariant four-velocity of the fluid elements, $k_\rho$ denotes the photon four-momentum, and $B_\sigma$ represents the comoving magnetic field four vector.

To investigate how the nonsingular ABG spacetime geometry imprints on the polarimetric observables, we consider following two distinct physical scenarios for the underlying magnetic field configuration around the black hole:
\begin{enumerate}
	\item[\rm (1)] The first magnetic field configuration adopted in this work follows the setup of Refs.\cite{Wang:2025btn,Hou:2023bep,Ruffini:1975ne}.
This magnetic field satisfies the ideal magnetohydrodynamic condition $F_{\mu\nu}u^\nu = 0$, and the corresponding gauge potential is axisymmetric and stationary. 
With a vanishing field line angular velocity, $\Omega_B = 0$, the magnetic four-vector field measured in the comoving frame is
	\begin{eqnarray}
		\label{eq:b_monopole}
		B^\mu &=& \frac{\partial_\theta A_\phi}{\rho^2 \sin\theta u^r} \left( u_t u^\mu + \delta^\mu_t \right)\nonumber\\
&=&\frac{\partial_\theta A_\phi}{\rho^2 \sin\theta u^r}\left(u_tu^t+1,\,u_tu^r,\,0,\,u_tu^\phi\right),
	\end{eqnarray}
	where the angular derivative of the gauge potential adopts a split monopole configuration \cite{Blandford:1977ds}
	\begin{equation}
		\label{eq:A_phi_derivative}
		\partial_\theta A_\phi = \Phi_0 \text{sign}(\cos\theta)\sin\theta,
	\end{equation}
and $\Phi_0$ is a constant characterizing magnetic flux.

	\item[\rm (2)] The second magnetic field configuration adopted in this work is a pure numerical one, $(B_r, B_\theta, B_\phi) = (0.87, 0, 0.5)$ in the plasma rest frame. This purely equatorial magnetic field is employed in a semi-analytic thin-ring model to calibrate to reproduce the polarization features of M87 \cite{Gelles:2021,Palumbo2020}.

\end{enumerate}

The polarization four-vector is parallel-transported along a null geodesic
\begin{equation}
	\label{eq:parallel_trans_eq}
	\frac{d f^\mu}{d \lambda} + \Gamma^\mu_{\ \nu\rho} f^\nu k^\rho = 0.
\end{equation}
The Stokes parameters are convenient physical quantities for characterizing the polarization properties of light, and are widely used in radiative transfer and polarimetric observations.
To avoid confusion with the ABG parameter $Q$, we denote the linear polarization Stokes parameters by $\mathcal{Q}$ and $\mathcal{U}$. After projecting the transported polarization onto the observer's screen, we define normalized transverse components
\begin{equation}
	\hat f_{X,n}=\frac{f_{X,n}}{\sqrt{f_{X,n}^2+f_{Y,n}^2}},\qquad
	\hat f_{Y,n}=\frac{f_{Y,n}}{\sqrt{f_{X,n}^2+f_{Y,n}^2}}.
\end{equation}
The Stokes parameters corresponding to this polarization are given by
\begin{equation}
\label{eq:stokes_components}
	\mathcal{Q}_n = I_n (\hat f_{X,n}^2 - \hat f_{Y,n}^2), \quad
	\mathcal{U}_n = 2 I_n \hat f_{X,n}\hat f_{Y,n},
\end{equation}
where $I_n=f_ng_n^3j(r_n)$. Truncating the sum over crossings $n$ at $N_{\text{max}}=3$, the total Stokes parameters are obtained by incoherently summing the contribution from each $n$: 
\begin{equation}	
\label{eq:stokes_sum}
	\mathcal{Q} = \sum_{n=1}^{N_{\text{max}}} \mathcal{Q}_n, \quad
	\mathcal{U} = \sum_{n=1}^{N_{\text{max}}} \mathcal{U}_n.
\end{equation}
The polarization intensity $P_{\text{obs}}$ and the $\text{EVPA}$ are given by
\begin{equation}
\label{eq:P_EVPA_def}
	P_{\text{obs}} = \sqrt{\mathcal{Q}^2 + \mathcal{U}^2}, \quad
	\text{EVPA} = \frac{1}{2}\operatorname{atan2}(\mathcal{U},\mathcal{Q}).
\end{equation}			
The degree of polarization is defined as:
\begin{equation}
m = P_{\text{obs}}/I_{\text{obs}},
\end{equation}
which equals to one for linearly polarized light. 
Lensing-induced polarization mixing leads to depolarization of the light on the screen. Let's consider two incoherent and linearly polarized light rays with Stokes parameters $(\mathcal{Q}_i, \mathcal{U}_i)$ and EVPAs $\chi_i,(i=1,2)$. The Stokes parameters of their incoherent superposition are 
 $\mathcal{Q}=\mathcal{Q}_1+\mathcal{Q}_2$ and $\mathcal{U}=\mathcal{U}_1+\mathcal{U}_2$, then we have
\begin{align}
\label{eq:depol_proof}
	P_{\text{obs}}^2 &= (\mathcal{Q}_1 + \mathcal{Q}_2)^2 + (\mathcal{U}_1 + \mathcal{U}_2)^2 \nonumber \\
			     &= P_1^2 + P_2^2 + 2(\mathcal{Q}_1 \mathcal{Q}_2 + \mathcal{U}_1 \mathcal{U}_2) \nonumber \\
		       	&= P_1^2 + P_2^2 + 2 P_1 P_2 \cos\left[2(\chi_1 - \chi_2)\right].
\end{align}
If $\chi_1 =\chi_2$, then $P_{\text{obs}}=P_1+P_2=I_1+I_2=I_{\text{obs}}$, and we have $m=1$, i.e., the sum of two beams of linearly polarized light with the same polarization direction remains linearly polarized. If $\chi_1 \neq\chi_2$ and defining $\eta=I_2/I_1$, then Eq.~\eqref{eq:depol_proof} yields
\begin{equation}
\label{eq:m_less_one}
	m^2
	=1-\frac{4\eta}{(1+\eta)^2}\sin^2(\chi_1-\chi_2).
\end{equation}
Thus, for strong depolarization to occur, the two light rays must have comparable amplitudes and a large EVPA separation. Maximum depolarization occurs for $\eta=1$ and $|\chi_1-\chi_2|=\pi/2$.

\section{Numerical Results: Optical and Polarized Images}
\label{sec:results}

Because the optical and polarimetric appearances are observer-dependent, we first specify the observer. In this work,  we consider a zero-angular-momentum observer (ZAMO) located at $r_{\text{obs}}=1000M$ with an inclination angle $\theta_{\text{obs}}$ \cite{Pu:2016eml}. 
An orthonormal tetrad basis $e_{(\mu)}$ associated with the ZAMO frame are chosen as follows:
\begin{align}
	e_{(0)} &= \xi(1, 0, 0, -\gamma), \\
	e_{(1)} &= \left(0, -\frac{1}{\sqrt{g_{rr}}}, 0, 0\right), \\
	e_{(2)} &= \left(0, 0, \frac{1}{\sqrt{g_{\theta\theta}}}, 0\right), \\
	e_{(3)} &= \left(0, 0, 0, -\frac{1}{\sqrt{g_{\phi\phi}}}\right),
\end{align}
where the function $\xi$ and the angular velocity parameter $\gamma$ are defined by
\begin{equation}
	\xi = \sqrt{\frac{-g_{\phi\phi}}{g_{tt}g_{\phi\phi} - g_{t\phi}^2}}, \qquad \gamma = \frac{g_{t\phi}}{g_{\phi\phi}}.
\end{equation}
The imaging apparatus is modeled as a panoramic wide-angle fisheye lens. 
The local momentum projected onto the ZAMO tetrad for a light ray with four-momentum $k^\nu$ are given by $p^{(\mu)} =e^{(\mu)}{}_\nu k^\nu $.
Each individual null geodesic entering the camera aperture is uniquely indexed by the celestial coordinates $\Theta$ and $\Psi$, which satisfy
\begin{equation}
	\cos \Theta = \frac{p^{(1)}}{p^{(0)}}, \qquad \tan \Psi = \frac{p^{(3)}}{p^{(2)}}.
\end{equation}
To achieve proper visualization, these celestial coordinates are further mapped onto a two-dimensional Cartesian observation screen $(X, Y)$ utilizing the stereographic projection technique (see Ref.\cite{Hu:2020usx} for more detail about the coordinates):
\begin{equation}
	X = -2 \tan \frac{\Theta}{2} \sin \Psi, \qquad Y = -2 \tan \frac{\Theta}{2} \cos \Psi.
\end{equation}
The positive $X$-axis is directed along $-e_{(3)}$, and the positive $Y$-axis points along $-e_{(2)}$. 
The polarization vector $f^\mu$, parallel-transported along a null geodesic, is projected onto the observation screen. Its screen components $(f_X, f_Y)$ are given by
\begin{equation}
	f_X = -e_{(3)\mu} f^\mu, \qquad f_Y = -e_{(2)\mu}f^\mu .
\end{equation}

\begin{figure*}[tb]
	\centering
	\begin{subfigure}[t]{0.24\textwidth}
		\centering
		\includegraphics[width=\textwidth]{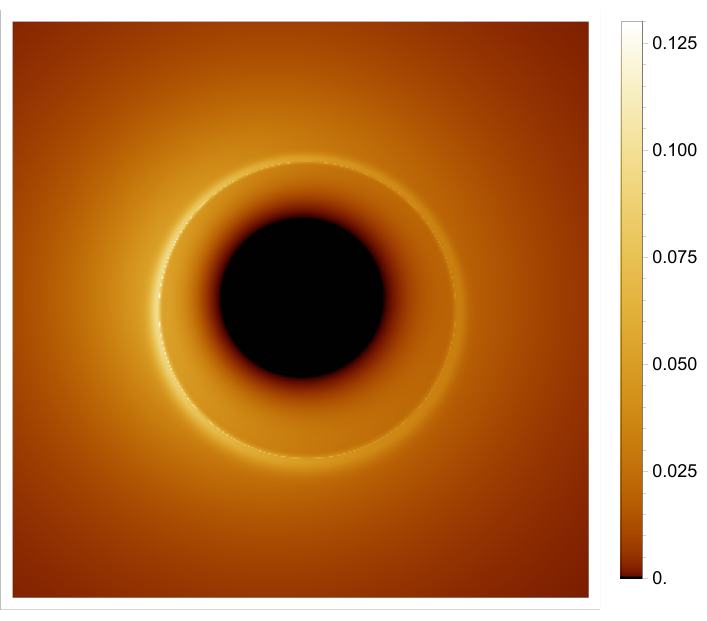}
		\caption{$a=0.4, Q=0$}
	\end{subfigure}
	\begin{subfigure}[t]{0.24\textwidth}
		\centering
		\includegraphics[width=\textwidth]{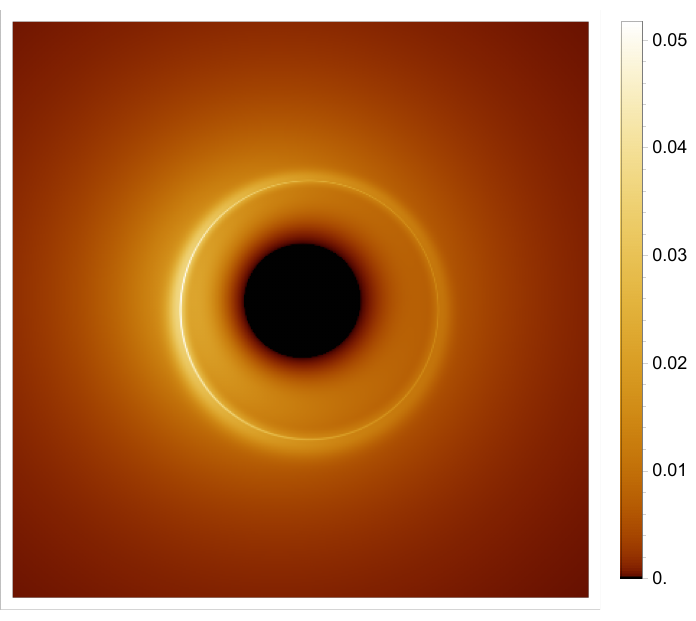}
		\caption{$a=0.4, Q=0.55$}
	\end{subfigure}
	\begin{subfigure}[t]{0.24\textwidth}
		\centering
		\includegraphics[width=\textwidth]{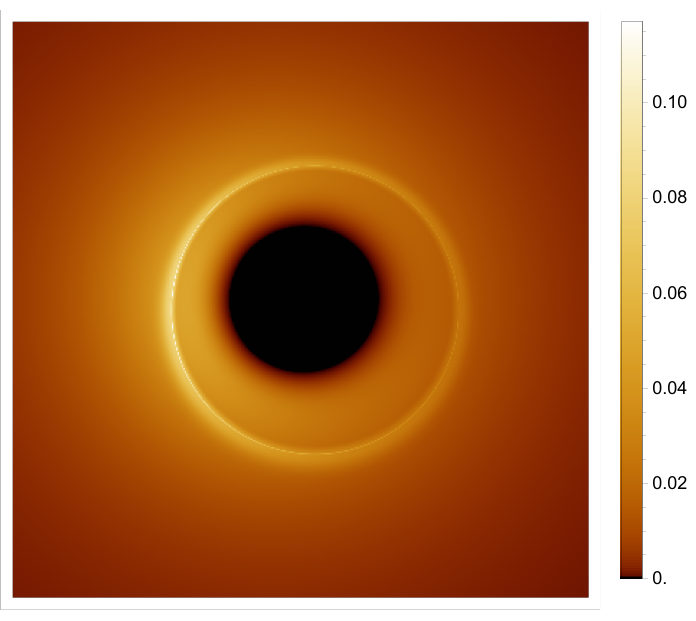}
		\caption{$a=0.8, Q=0$}
	\end{subfigure}
	\begin{subfigure}[t]{0.24\textwidth}
		\centering
		\includegraphics[width=\textwidth]{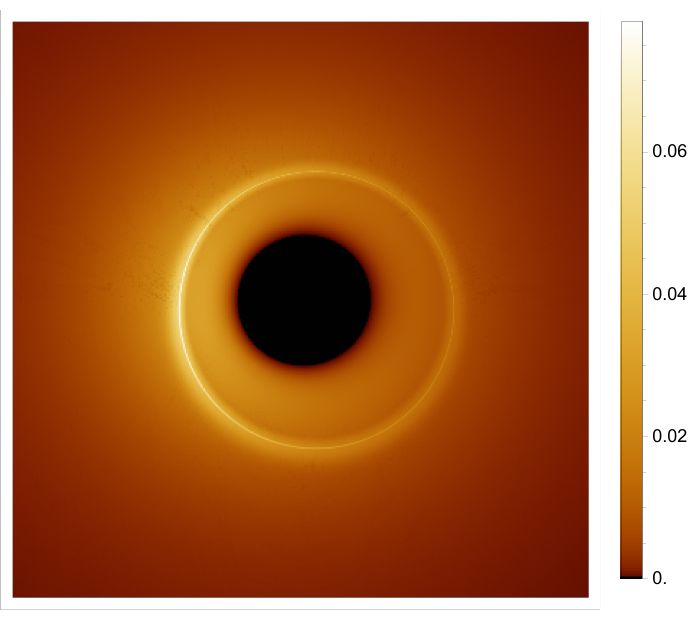}
		\caption{$a=0.8, Q=0.3$}
	\end{subfigure}
	
	\vspace{0.1cm}
	
	\begin{subfigure}[t]{0.24\textwidth}
		\centering
		\includegraphics[width=\textwidth]{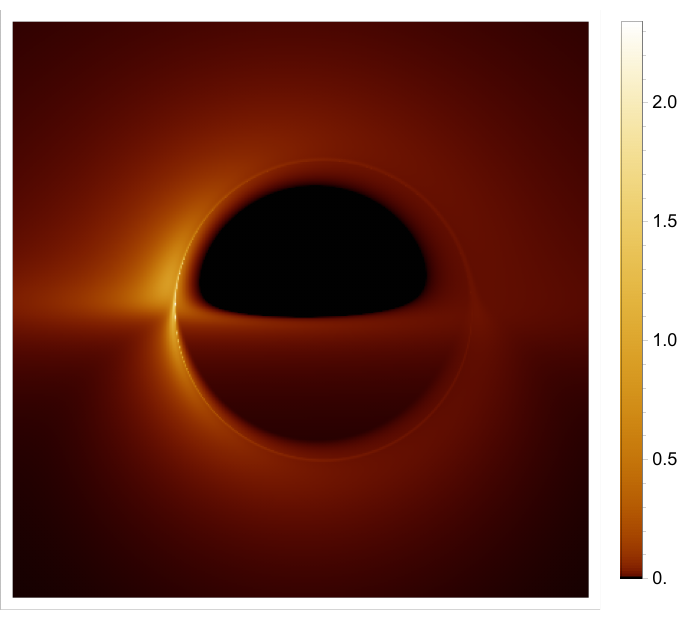}
		\caption{$a=0.4, Q=0$}
	\end{subfigure}
	\begin{subfigure}[t]{0.24\textwidth}
		\centering
		\includegraphics[width=\textwidth]{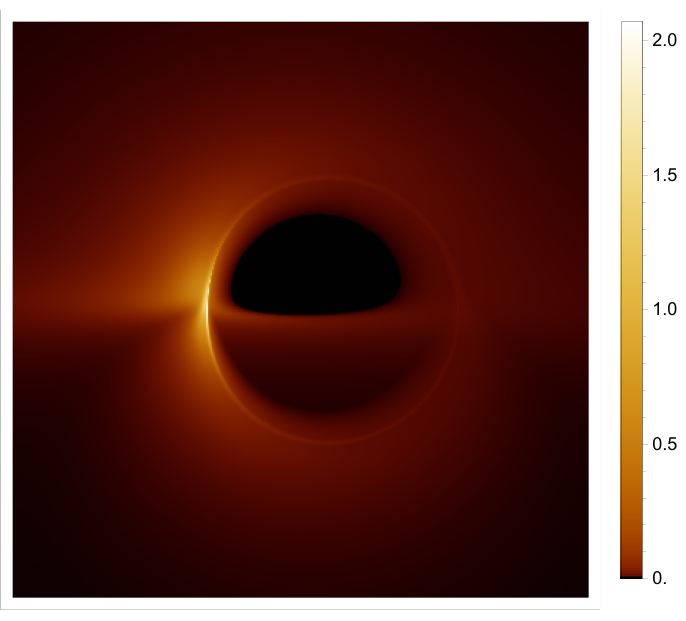}
		\caption{$a=0.4, Q=0.55$}
	\end{subfigure}
	\begin{subfigure}[t]{0.24\textwidth}
		\centering
		\includegraphics[width=\textwidth]{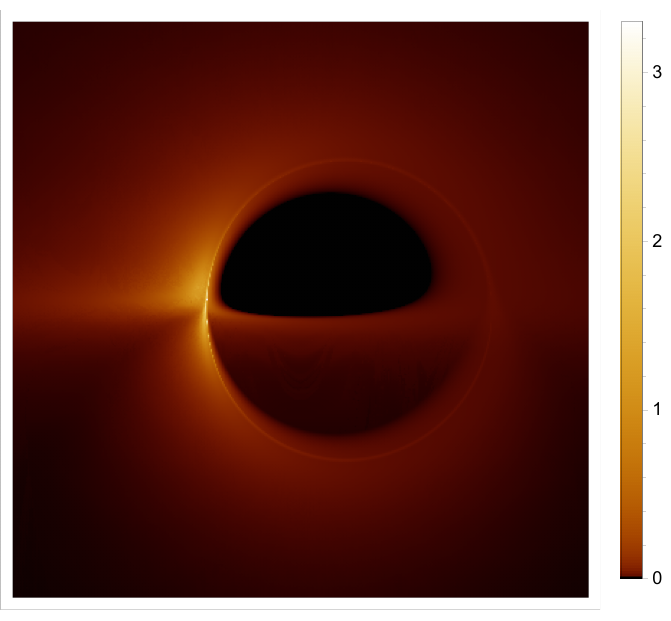}
		\caption{$a=0.8, Q=0$}
	\end{subfigure}
	\begin{subfigure}[t]{0.24\textwidth}
		\centering
		\includegraphics[width=\textwidth]{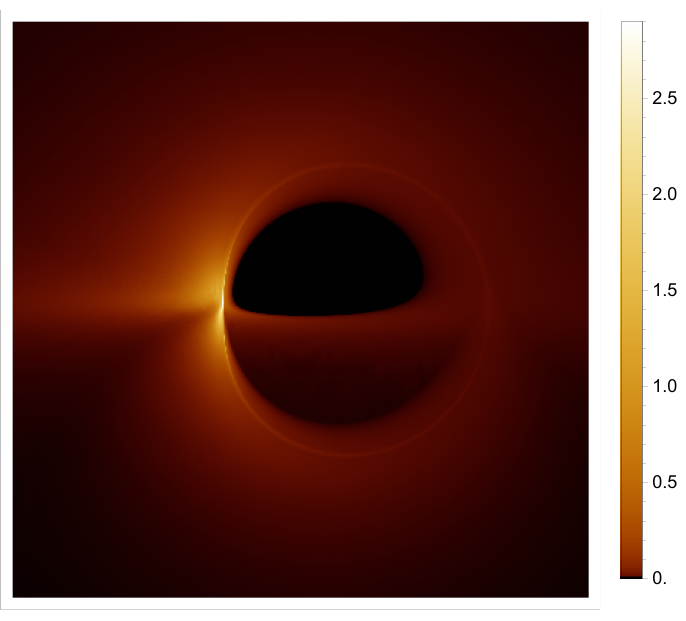}
		\caption{$a=0.8, Q=0.3$}
	\end{subfigure}
	\caption{Observed intensity images of rotating ABG black holes. The horizontal and vertical axes represent the impact parameters $X/M$ and $Y/M$, respectively. The top row corresponds to an inclination of $\theta_{\rm obs} = 17^\circ$, and the bottom row to $\theta_{\rm obs} = 83^\circ$.}
	\label{fig:appearance}
\end{figure*}

Based on the ZAMO observer described above, we obtain the optical appearance of the rotating ABG geometry for various parameters, as illustrated in Fig.~\ref{fig:appearance}. We consider cases of slow rotation ($a=0.4$) and fast rotation ($a=0.8$), and compare the Kerr limit ($Q=0$) cases with the rotating ABG cases to see the impact of the regularization parameter $Q$ on the images.
The observation inclination angles are set to $17^\circ$ (motivated by M87*) and $83^\circ$ (high inclination). 
Images in the upper row are observed at an inclination angle of $17^\circ$. In each image, there is a prominent inner shadow and a bright photon ring, which is associated with the strongly lensed effect near the photon sphere. 
From the images in the lower row observed at an inclination angle of $83^\circ$, we can see that compared with the slow-rotation cases ($a=0.4$), the bright photon rings of the fast-rotation cases ($a=0.8$) exhibit characteristic D-shaped morphology, which is consistent with stronger frame-dragging effect in the fast-rotation cases. 
For all the cases with our chosen parameters, increasing $Q$ contracts both the bright photon rings and the inner shadows. It is also observed that the size of the inner shadow changes more strongly than that of the photon ring, which suggests that the inner shadow could be a more sensitive geometric diagnostic to test the rotating ABG model. 
 At high inclination angles, nonzero $Q$ also affects the D-shaped distortion.

\begin{figure*}[tb]
	\centering
	\begin{subfigure}[t]{0.24\textwidth}
		\centering
		\includegraphics[width=\textwidth]{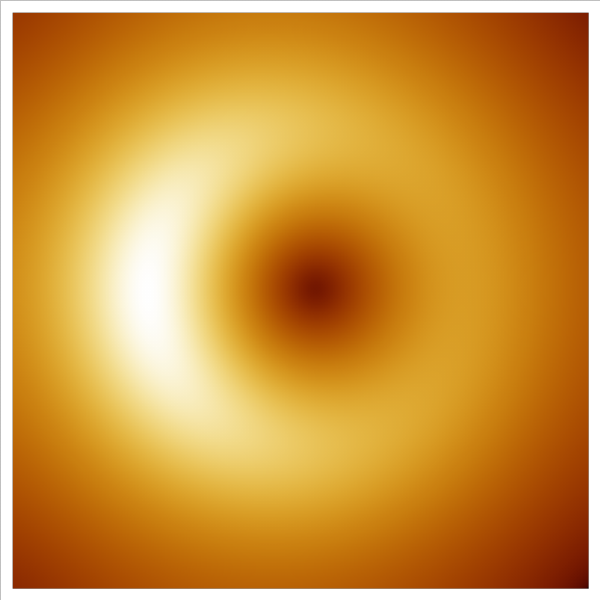}
		\caption{$a=0.4, Q=0$}
	\end{subfigure}
	\begin{subfigure}[t]{0.24\textwidth}
		\centering
		\includegraphics[width=\textwidth]{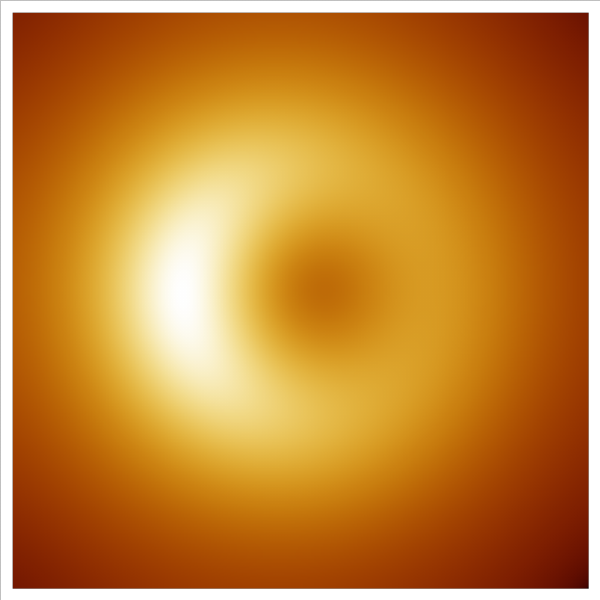}
		\caption{$a=0.4, Q=0.55$}
	\end{subfigure}
	\begin{subfigure}[t]{0.24\textwidth}
		\centering
		\includegraphics[width=\textwidth]{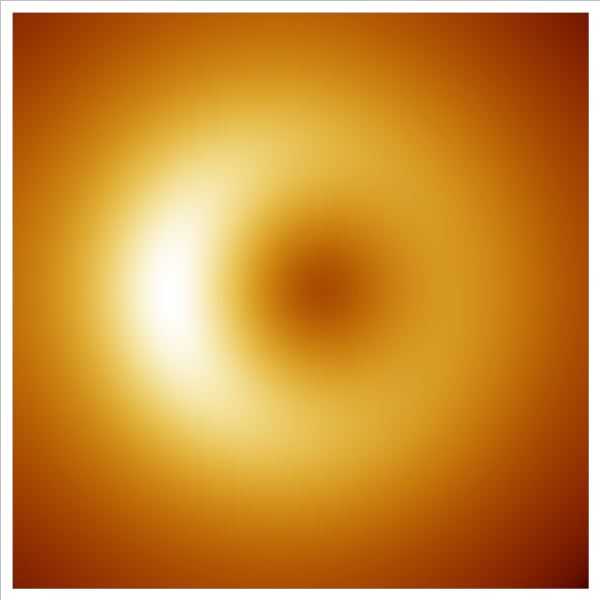}
		\caption{$a=0.8, Q=0$}
	\end{subfigure}
	\begin{subfigure}[t]{0.24\textwidth}
		\centering
		\includegraphics[width=\textwidth]{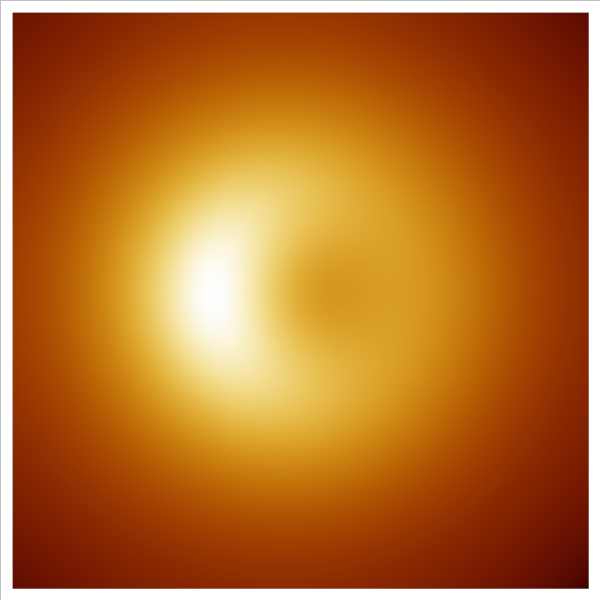}
		\caption{$a=0.8, Q=0.3$}
	\end{subfigure}
	
	\vspace{0.1cm}
	
	\begin{subfigure}[t]{0.24\textwidth}
		\centering
		\includegraphics[width=\textwidth]{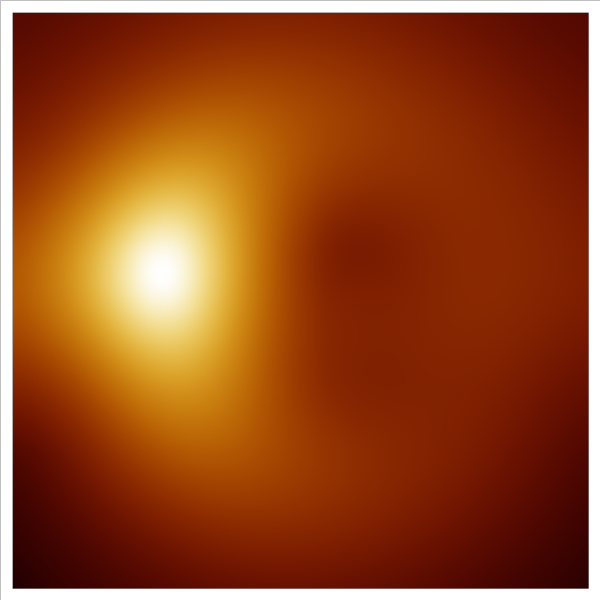}
		\caption{$a=0.4, Q=0$}
	\end{subfigure}
	\begin{subfigure}[t]{0.24\textwidth}
		\centering
		\includegraphics[width=\textwidth]{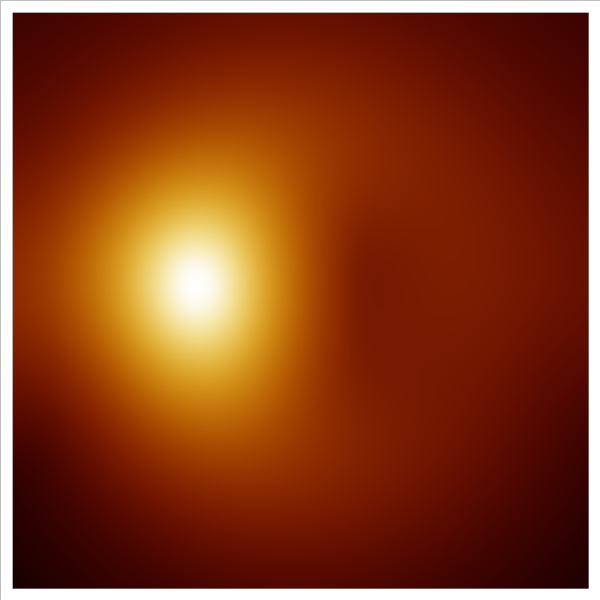}
		\caption{$a=0.4, Q=0.55$}
	\end{subfigure}
	\begin{subfigure}[t]{0.24\textwidth}
		\centering
		\includegraphics[width=\textwidth]{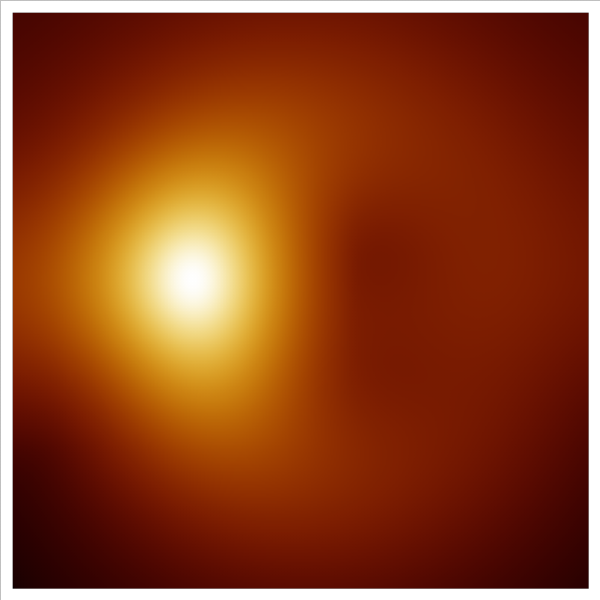}
		\caption{$a=0.8, Q=0$}
	\end{subfigure}
	\begin{subfigure}[t]{0.24\textwidth}
		\centering
		\includegraphics[width=\textwidth]{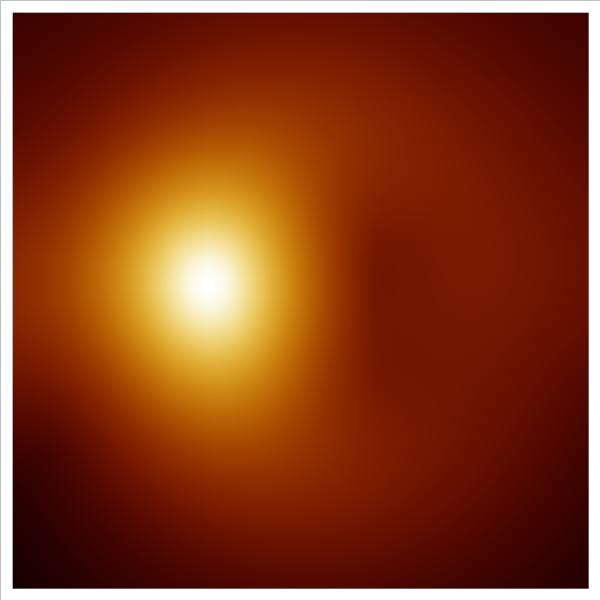}
		\caption{$a=0.8, Q=0.3$}
	\end{subfigure}
	\caption{Gaussian-blurred intensity images of rotating ABG black holes. The horizontal and vertical axes represent the impact parameters $X/M$ and $Y/M$, respectively. The top row corresponds to an inclination of $\theta_{\rm obs} = 17^\circ$, and the bottom row to $\theta_{\rm obs} = 83^\circ$.}
	\label{fig:blurred_images}
\end{figure*}

The preceding discussion highlights differences between the optical images at ideal resolution of Kerr black holes and that of rotating ABG geometries. 
To mimic the finite instrumental resolution,  we convolve each image with a Gaussian kernel whose width is $1/12$ of the field of view (FOV). The resulting images are presented in Fig.~\ref{fig:blurred_images}. This operation is not a complete EHT measurement simulation: it does not include an angular mass-distance conversion, sparse interferometric coverage, thermal noise, calibration uncertainties, or image reconstruction. For images observed at an inclination angle of $17^\circ$, the central brightness depression in the nonzero-$Q$ rotating ABG model is smaller and shallower than that in their Kerr counterparts.  In the case with $a=0.8$ and $Q=0.3$, the central brightness depression nearly merges with the surrounding bright region.
After the convolution, the images observed at an inclination angle of $83^\circ$ exhibit weaker visual dependence on model parameters than those at $17^\circ$.
 These image-domain differences cannot confirm detectability with the current EHT, but motivate future high-resolution observations.

\begin{figure*}[tb]
	\centering
	\begin{subfigure}[t]{0.24\textwidth}
		\centering
		\includegraphics[width=\textwidth]{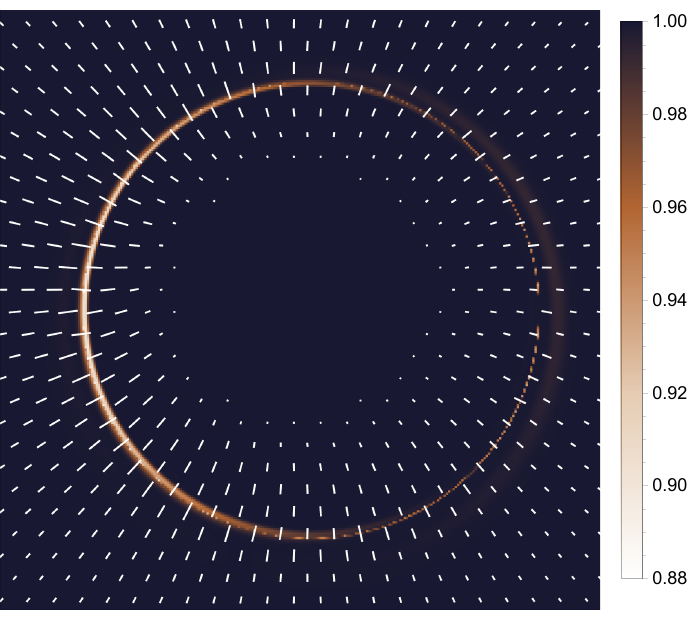}
		\caption{$a=0.4, Q=0$}
	\end{subfigure}
	\begin{subfigure}[t]{0.24\textwidth}
		\centering
		\includegraphics[width=\textwidth]{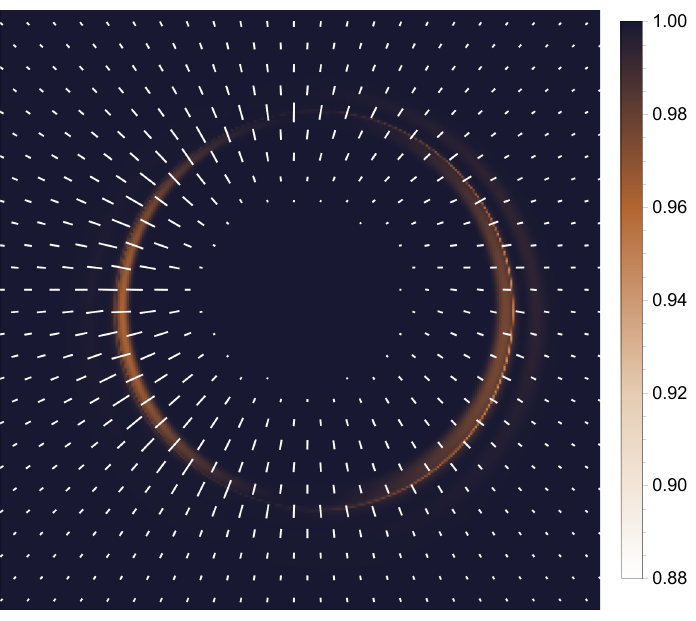}
		\caption{$a=0.4, Q=0.55$}
	\end{subfigure}
	\begin{subfigure}[t]{0.24\textwidth}
		\centering
		\includegraphics[width=\textwidth]{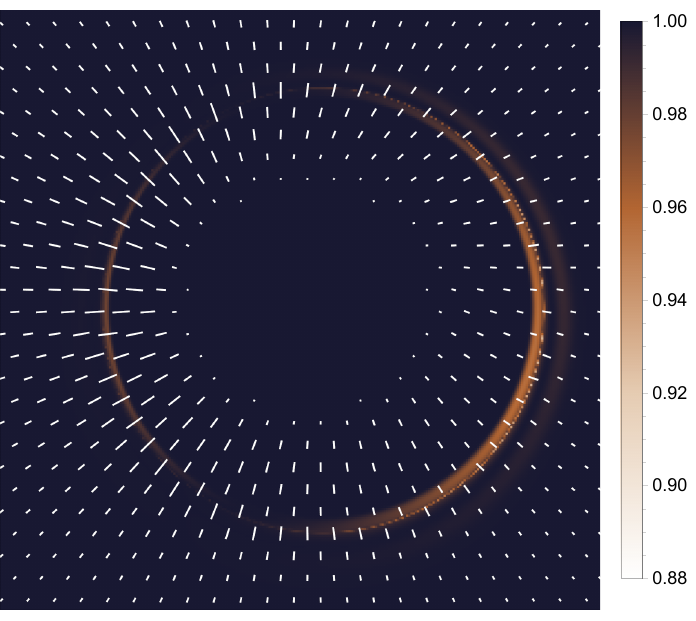}
		\caption{$a=0.8, Q=0$}
	\end{subfigure}
	\begin{subfigure}[t]{0.24\textwidth}
		\centering
		\includegraphics[width=\textwidth]{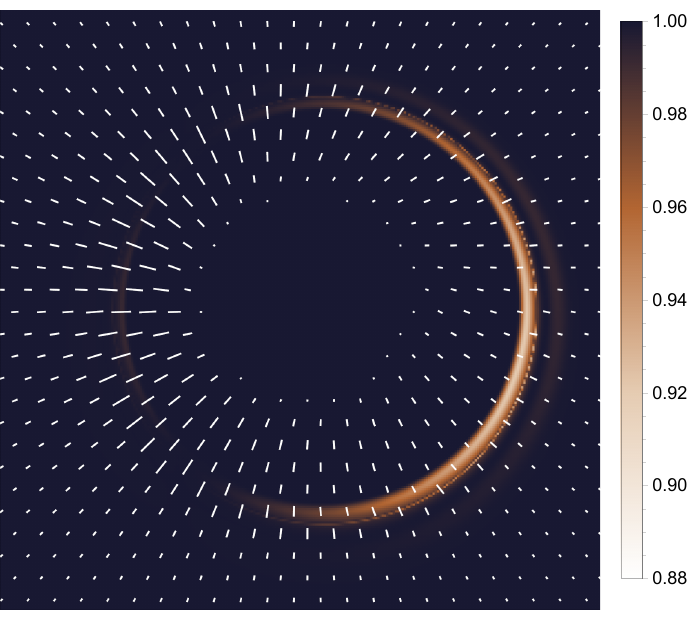}
		\caption{$a=0.8, Q=0.3$}
	\end{subfigure}
	
	\vspace{0.1cm}
	
	\begin{subfigure}[t]{0.24\textwidth}
		\centering
		\includegraphics[width=\textwidth]{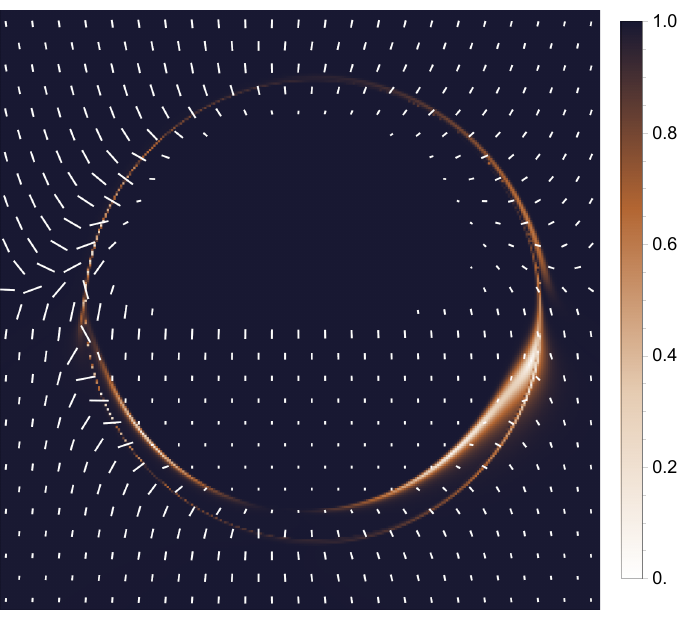}
		\caption{$a=0.4, Q=0$}
	\end{subfigure}
	\begin{subfigure}[t]{0.24\textwidth}
		\centering
		\includegraphics[width=\textwidth]{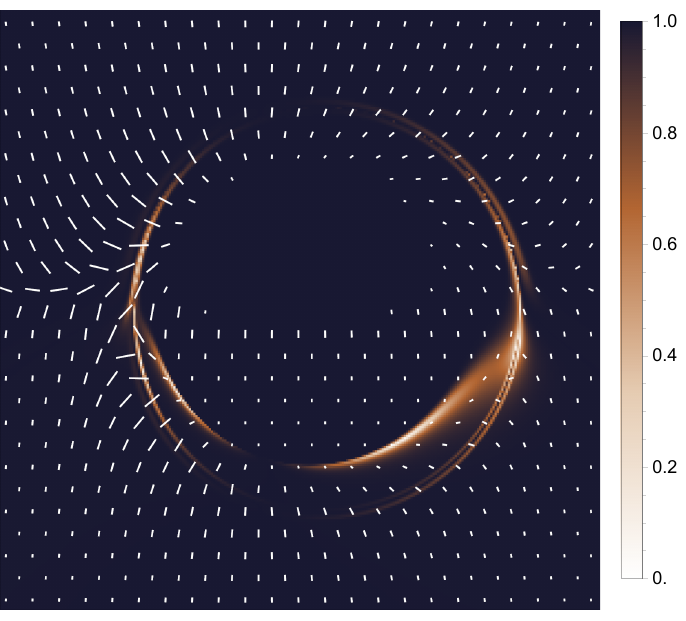}
		\caption{$a=0.4, Q=0.55$}
	\end{subfigure}
	\begin{subfigure}[t]{0.24\textwidth}
		\centering
		\includegraphics[width=\textwidth]{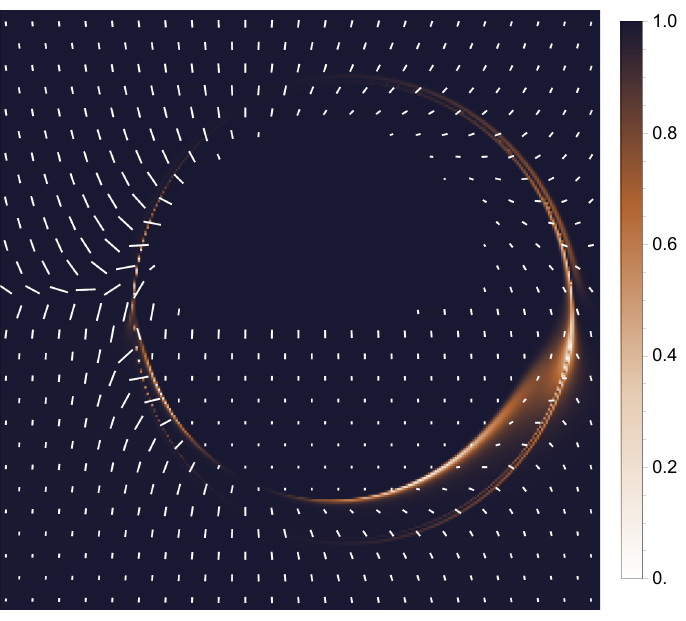}
		\caption{$a=0.8, Q=0$}
	\end{subfigure}
	\begin{subfigure}[t]{0.24\textwidth}
		\centering
		\includegraphics[width=\textwidth]{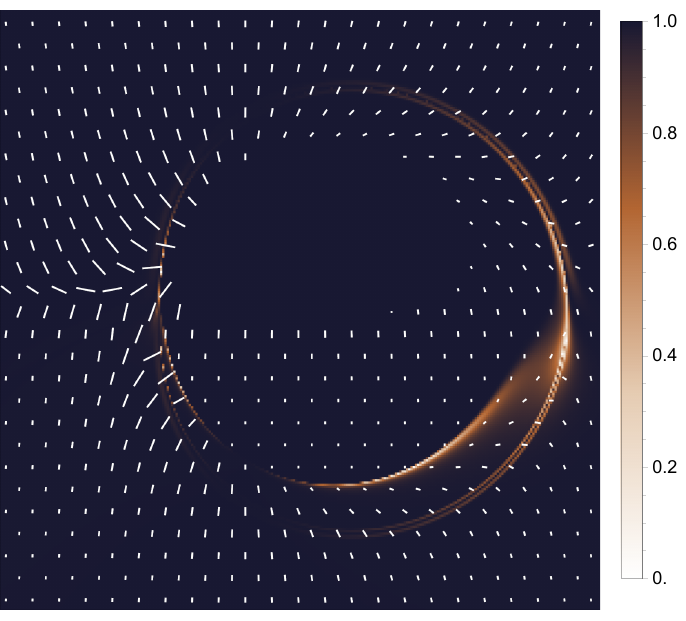}
		\caption{$a=0.8, Q=0.3$}
	\end{subfigure}
	\caption{Polarized images of rotating ABG black holes under the first magnetic field configuration. The horizontal and vertical axes represent the impact parameters $X/M$ and $Y/M$, respectively. The top row corresponds to an inclination of $\theta_{\rm obs} = 17^\circ$, and the bottom row to $\theta_{\rm obs} = 83^\circ$.}
	\label{fig:polarization}
\end{figure*}

\begin{figure*}[tb]
	\centering
	\begin{subfigure}[t]{0.24\textwidth}
		\centering
		\includegraphics[width=\textwidth]{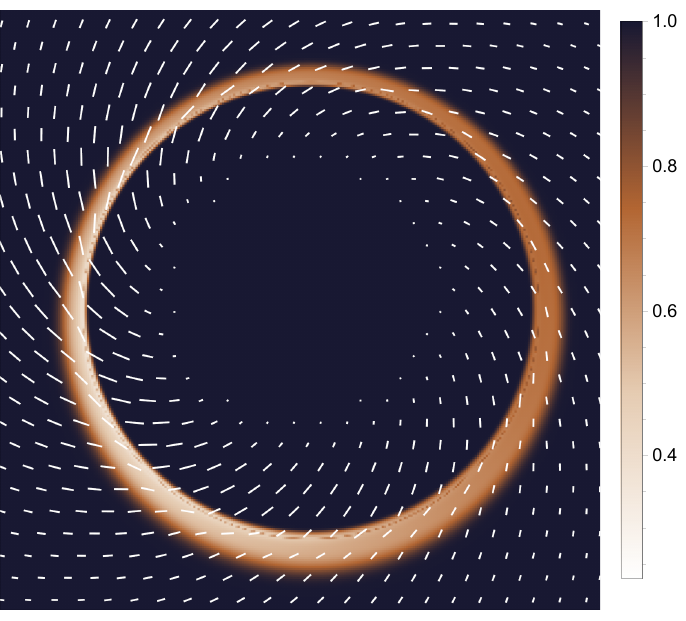}
		\caption{$a=0.4, Q=0$}
	\end{subfigure}
	\begin{subfigure}[t]{0.24\textwidth}
		\centering
		\includegraphics[width=\textwidth]{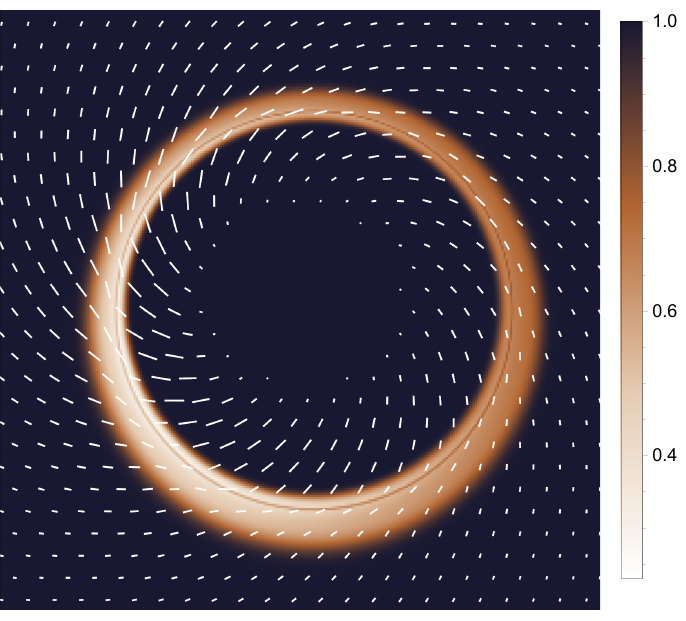}
		\caption{$a=0.4, Q=0.55$}
	\end{subfigure}
	\begin{subfigure}[t]{0.24\textwidth}
		\centering
		\includegraphics[width=\textwidth]{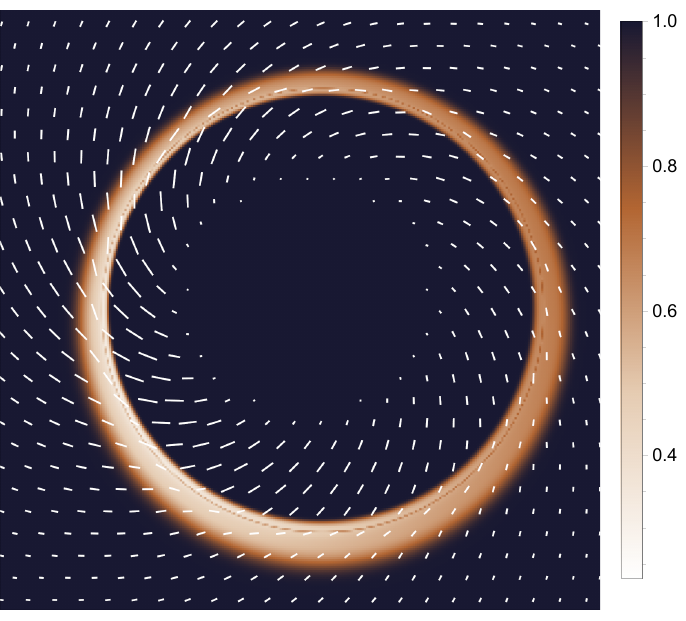}
		\caption{$a=0.8, Q=0$}
	\end{subfigure}
	\begin{subfigure}[t]{0.24\textwidth}
		\centering
		\includegraphics[width=\textwidth]{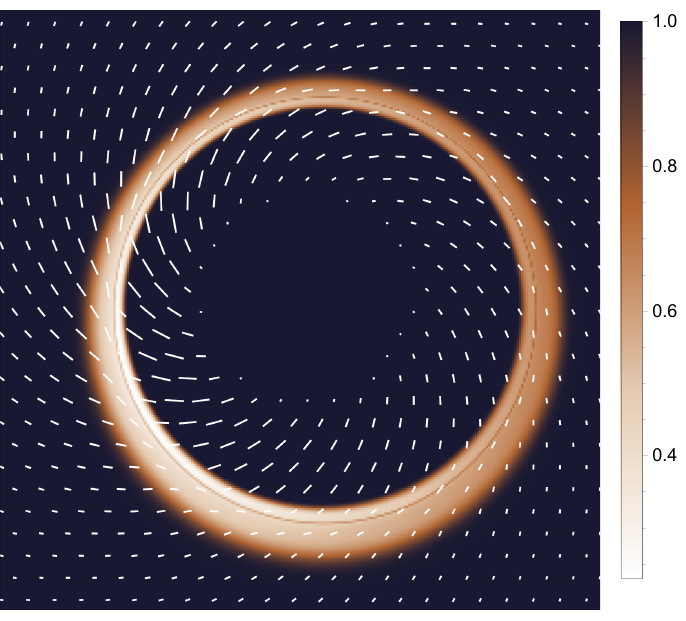}
		\caption{$a=0.8, Q=0.3$}
	\end{subfigure}
	
	\vspace{0.1cm}
	
	\begin{subfigure}[t]{0.24\textwidth}
		\centering
		\includegraphics[width=\textwidth]{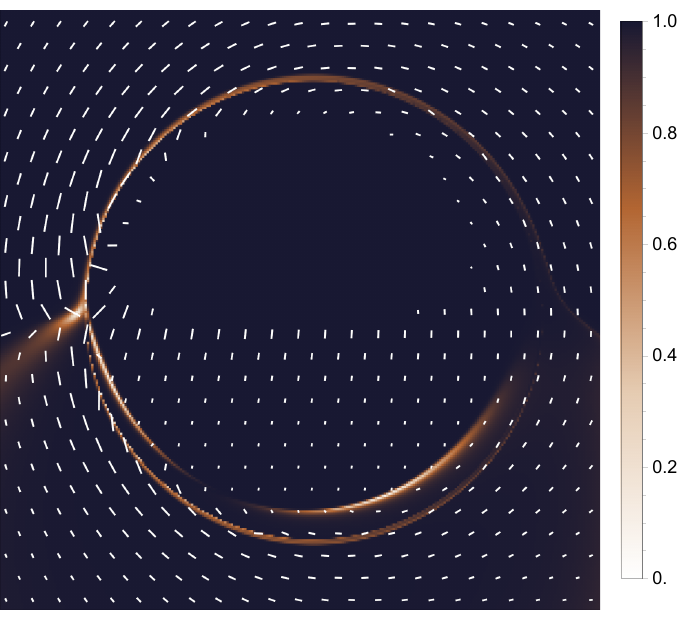}
		\caption{$a=0.4, Q=0$}
	\end{subfigure}
	\begin{subfigure}[t]{0.24\textwidth}
		\centering
		\includegraphics[width=\textwidth]{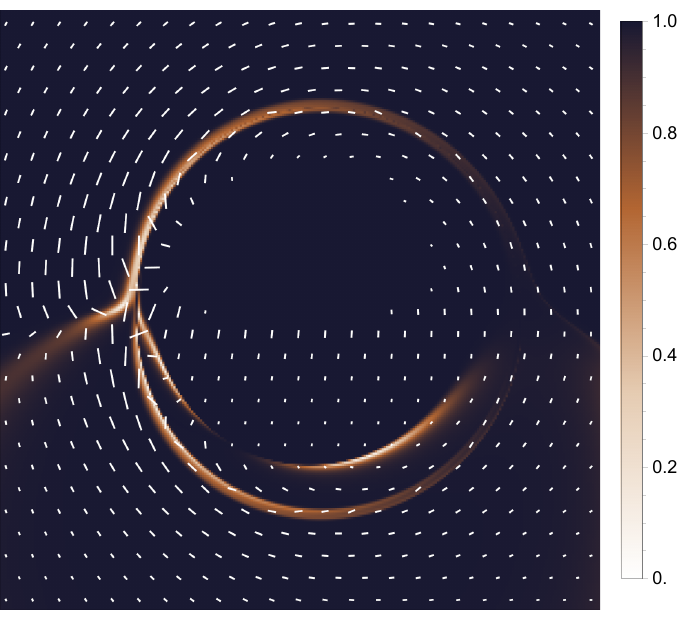}
		\caption{$a=0.4, Q=0.55$}
	\end{subfigure}
	\begin{subfigure}[t]{0.24\textwidth}
		\centering
		\includegraphics[width=\textwidth]{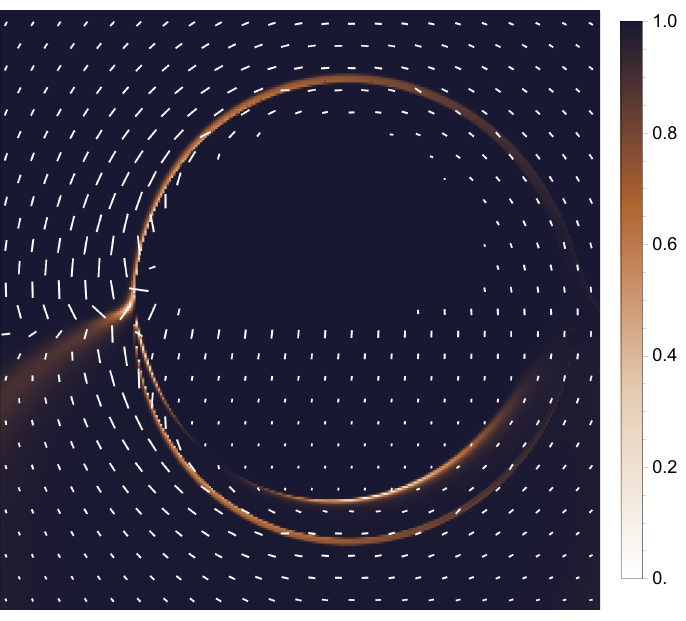}
		\caption{$a=0.8, Q=0$}
	\end{subfigure}
	\begin{subfigure}[t]{0.24\textwidth}
		\centering
		\includegraphics[width=\textwidth]{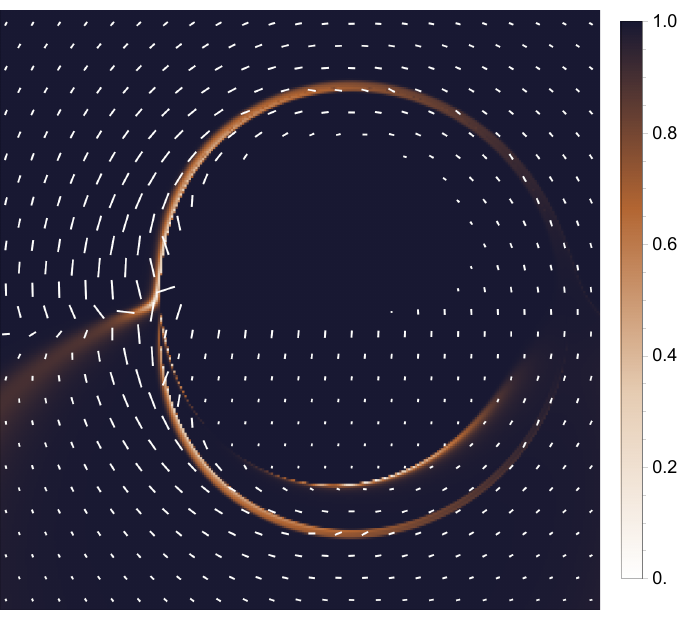}
		\caption{$a=0.8, Q=0.3$}
	\end{subfigure}
	\caption{Polarized images of rotating ABG black holes under the second magnetic field configuration. The horizontal and vertical axes represent the impact parameters $X/M$ and $Y/M$, respectively. The top row corresponds to an inclination of $\theta_{\rm obs} = 17^\circ$, and the bottom row to $\theta_{\rm obs} = 83^\circ$.}
	\label{fig:polarization1}
\end{figure*}

In Figs.~\ref{fig:polarization} and \ref{fig:polarization1}, we present the polarized images for the two magnetic field configurations mentioned in previous section. 
The background color shows the degree of polarization $m$; lighter color indicates stronger cancellation among the Stokes contributions from different equatorial disk crossings. 
The superimposed white ticks show the screen-projected EVPA pattern, and the length of a tick is proportional to the polarization intensity $P_{\text{obs}}$. Because linear polarization is invariant under a $180^\circ$ rotation, the ticks have no arrowheads. The EVPAs are uniformly subsampled for visual clarity.

Fig.~\ref{fig:polarization} shows the observed polarized images of the rotating ABG black hole under the first magnetic field configuration.
The corresponding Kerr cases with $Q=0$ are also shown for comparison. Since the emitted light from the accretion disk is supposed to be linearly polarized, 
a pixel on the image plane receiving only one disk-crossing contribution has degree of polarization $m=1$, while a pixel receiving lensing-induced multiple disk-crossing contributions has lower degree of polarization due to the polarization mixing and the depolarization effect for the superimposed light. 
The regularization parameter $Q$ has a negligible effect on the EVPA pattern, yet exerts a pronounced influence on the distribution of the degree of polarization.

The images in the upper row of Fig.~\ref{fig:polarization} show the polarized images observed at an inclination angle of $\theta_{\text{obs}}=17^\circ$. A ring-like depolarization region locates around the central inner shadow and this depolarization region almost coincides with the bright photon ring in the corresponding density image in Fig.\ref{fig:appearance}. The physical explanation for this observation is straightforward: 
the depolarization region arises from the mixing between lensing-induced multiple disk-crossing contributions, and for a low inclination angle, the lensing-induced twice and more disk-crossing contributions appear in the vicinity of the photon ring on the observation plane. 
 However, as demonstrated in Eq.~\eqref{eq:m_less_one}, multiple polarization mixing is necessary but not sufficient for strong depolarization, and additional requirements consist of comparable intensities and large EVPA separation between different contributions.
 So, the degree of polarization varies significantly along the ring-like region. 
For small spin parameter $a$ ($a=0.4$), the ring-like region exhibits approximate left–right symmetry; for large $a$ ($a=0.8$), this symmetry is strongly broken and the ring becomes highly asymmetric. For each spin value, the regularization parameter $Q$ reduces the radius of the ring-like depolarization region to a value noticeably smaller than that of the Kerr black hole. And, $Q$ also has an obvious impact on the distribution of the degree of polarization along the ring-like region.

The images in the lower row of Fig.~\ref{fig:polarization} show the polarized images observed at a large inclination angle ($\theta_{\text{obs}}=83^\circ$). An interesting feature appears that the ring-like low-polarization region splits into two distinct branches in its lower part. This feature is absent in the density image, making it a distinctive signature of the polarization image. A qualitative explanation for this multiple-branch feature is: for large inclination angle, the lower part of the secondary image exhibits substantial spatial spreading, so the spatial region over which a low degree of polarization may arise could become considerably extended; however, low degree of polarization further requires comparable intensities and large EVPA separation between the direct and secondary contributions, thus two branches of low-depolarization region appear. The degree of polarization also varies significantly along the ring-like region, and as the increase of the spin parameter $a$, the image exhibits stronger left–right asymmetry. The regularization parameter $Q$ also has an obvious impact on the radius of the ring-like low-polarization region and the distribution of the degree of polarization along the ring.

For the second magnetic configuration, $(B_r,B_\theta,B_\phi)=(0.87,0,0.5)$, the polarized images are shown in Fig.~\ref{fig:polarization1}. The upper row presents the images at an inclination angle of $17^\circ$ and the lower row corresponds to those at $83^\circ$. Although these images are qualitatively similar to those in Fig.~\ref{fig:polarization}, they still exhibit some differences in certain details. For each image in the upper row, the depolarization region changes from a thin ring-like morphology to a broader annular shape. 
The physical origin of this change lies in the variation of the magnetic field within the accretion disk since the geodesic trajectories and disk-crossing counting are unchanged.
Within the broad depolarization region, a high-polarization thin circle is observed. This indicates again the fact that multiple polarization mixing is necessary but not sufficient for strong depolarization. Compared with the Kerr black hole case, the regularization parameter $Q$ still has an obvious impact on the radius of the depolarization region.
For each image in the lower row with a larger inclination angle, the broad annular depolarization region becomes into a narrow ring-like region. Further, in the second magnetic configuration, this ring-like region exhibits three distinct branches in its lower part, in contrast to the two branches observed in the first magnetic configuration.  And, the regularization parameter $Q$ also has an obvious impact on the size of the ring and the distribution of the degree of polarization along the ring. The qualitative physical explanation for the features of the depolarization region is similar to that discussed in the first magnetic configuration case.
In order to obtain a quantitative understanding of the features, we need to track both the relative intensity of different contributions arriving at each screen point and the angular separation between their EVPAs.


\section{Conclusions}
\label{sec:Conclusion}

In this work, we consider two representative magnetic filed configurations and focus on the polarized images of the rotating ABG geometry surrounded by a stationary, geometrically thin, and optically thin equatorial accretion disk. 
General-relativistic backward ray tracing method and parallel transport of the polarization vectors allow us to numerically obtain the EVPA patterns, polarized
intensities and degree of polarization in the observation plane at a fixed viewing inclination angle. 
Our calculation treats photons as test fields following null geodesics of the background metric and is intended to explore how the observables are impacted by the model parameters, i.e., the spin $a$ and regularization parameter $Q$.
In order for comparison with the polarized images, intensity images for various model parameters are also presented, and illustrative Gaussian convolutions
of the intensity images show that the size of the central brightness depression are sensitive to the regularization parameter $Q$.

From the polarized images, it is found that the EVPA patterns show little sensitivity to the parameter $Q$ but are largely sensitive to the inclination angles of the observer. However, the morphologies of the low-polarization regions in the image planes and the distributions of the degree of polarization are sensitive to the regularization parameter $Q$.  
Further, at low inclination angles, the morphologies of the low-polarization regions are similar to that of the intensity images; at high inclination angles, the low-polarization regions have distinctive morphologies, and in particular, multiple-branch structures appear in the lower part of the low-polarization regions. The magnetic field configuration only affects the quantitative results, not the qualitative conclusions.

The results show that the intensity and polarimetric observables respond differently to the model parameters and motivate joint diagnostics for black hole studies using optical observations. The present work just demonstrates the influences of the regularization parameter on the optical images in simplified model setups. These influences are likely degenerate with those arising from more complex accretion disk models, magnetic field configurations and radiation transport processes.
For future works, the idealized equatorial disk can be generalized to a geometrically thick one, even with turbulent flows, and the transport of the polarized vectors can be extended to include Faraday rotation and conversion.

	\bibliography{refRABGbh}  
	
\end{document}